\documentclass{article}
\usepackage{amstex,psfig}

\bigskip
\begin{document}

\title{A Free Boundary Problem in the Theory of the Stars\thanks{Talk given by
S.S. Yazadjiev and T.L. Boyadjiev at the $4^{th}$ General Conference of the
Balkan Physical Union, Veliko Turnovo , Bulgaria, 22-25 August, 2000}}

\author{S. S. Yazadjiev\\
{\small \textit{Faculty of Physics, University of Sofia, Sofia,
Bulgaria}}\\
[-1.mm] {\small \textit{E-mail: yazad@@phys.uni-sofia.bg}}\\
\\
T. L. Boyadjiev\\ {\small \textit{Faculty of Mathematics and Computer Science}}\\
[-1.mm]{\small \textit{University of Sofia, Sofia, Bulgaria}}\\
[-1.mm]{\small \textit{E-mail: todorlb@@fmi.uni-sofia.bg}}\\\\
M. D. Todorov\\
{\small \textit{Faculty of Applied Mathematics and Computer
Science}}\\ [-1.mm]{\small\textit{Technical University of Sofia, Sofia, Bulgaria}}\\
[-1.mm]{\small\textit{E-mail: mtod@@vmei.acad.bg}} \\\\
P. P. Fiziev\\[-1.mm]{\small \textit{Faculty of Physics, University of Sofia, Sofia,
Bulgaria}}\\[-1.mm]{\small \textit{E-mail: fiziev@@phys.uni-sofia.bg}}}

\date{}
\maketitle

\begin{abstract}
We investigate numerically  models of the static spherically
symmetric boson-fermion stars in the scalar-tensor theory of
gravity with massive dilaton field. The proper mathematical model
of such stars is interpreted as a nonlinear two-parametric
eigenvalue problem with unknown internal boundary. To solve this
problem the Continuous Analogue of Newton Method is used.

\end{abstract}

\section{Main model}\label{sec2}
Boson stars are gravitationally bound macroscopic quantum states
made up of scalar bosons. They differ from the usual fermionic
stars in that they are only prevented from collapsing
gravitationally by the Heisenberg uncertainty principle. For a
self-interacting boson field the mass of the boson star, even for
small values of the coupling constant, turns out to be of the
order of Chandrasekhar's mass when the boson mass is similar to
proton mass. Thus, the boson stars arise as possible candidates
for non-baryonic dark matter in the universe and, consequently,
as a possible solution of one of the outstanding problems in
modern astrophysics: the problem of non luminous matter in the
universe. Most of the stars are of primordial origin, being
formed from an original gas of fermions and bosons in the early
universe. That is why it should be expected that most stars are a
mixture of both fermions and bosons in different proportions.

Boson-fermion stars are also a good model to understand more
about the nature of strong gravitational fields not only in
general relativity but also in the other theories of gravity.

The most natural and promising generalizations of general
relativity are the scalar-tensor theories of gravity \cite{DEF1}.
In these theories the gravity is mediated not only by a tensor
field (the metric of space-time) but also by a scalar field (the
dilaton). These dilatonic theories of gravity contain arbitrary
functions of the scalar field that determine the gravitational
``constant'' as a dynamical variable and the strength of the
coupling between the scalar field and matter. It should be
stressed that specific scalar-tensor theories of gravity arise
naturally as a low energy limit of the string theory \cite{GSW}
which is the most promising modern model of unification of all
fundamental physical interactions.

Boson stars in the scalar-tensor theories of gravity with
massless dilaton have been widely investigated recently both
numerically and analytically \cite{Y} (and references therein).
Mixed boson-fermion stars in scalar tensor theories of gravity
however have not been investigated so far in contrast to general
relativistic case where boson-fermion stars have been
investigated \cite{HLM}.

We will consider a static and spherically symmetric mixed
boson-fermion star in asymptotic flat space-time. Then the main
system of differential equations can be written in the following
form \cite{tomiplas}:
\begin{align}
 \frac{d\lambda }{dr}=&F_1\equiv \frac{1-\exp(\lambda
)}{r}+r\,\left\{ \exp (\lambda )\left[
\stackrel{F}{T_{0}^{0}}+\stackrel{B}{T_{0}^{0}}+\frac{1}{2}\gamma
^{2}V(\varphi )\right] +\left( \frac{d\varphi}{dr}\right)
^{2}\right\}, \nonumber \\ \frac{d\nu }{dr}=&F_2\equiv \!
-\frac{1-\exp (\lambda )}{r}-r\,\left\{\exp (\lambda )\left[
\stackrel{F}{T_{1}^{1}}+\stackrel{B}{T_{1}^{1}}+\frac{1}{2}\gamma
^{2}V(\varphi )\right]-\left( \frac{d\varphi }{dr}\right)
^{2}\right\}, \nonumber \\ \frac{d^{2}\varphi
}{dr^{2}}=&F_3\equiv -\frac{2}{r}\frac{d\varphi
}{dr}+\frac{1}{2}\left( F_{1}-F_{2}\right) \frac{d\varphi }{dr}
+\frac{1}{2}\exp (\lambda )\,\left[ \alpha (\varphi )\,(\stackrel{F}{T}+
\stackrel{B}{T})+\frac{1}{2}\gamma ^{2}V^{^{\prime }}(\varphi )\right] ,
\label{eqi} \\
\frac{d^{2}\sigma }{dr^{2}}=&F_4\equiv -\frac{2}{r}\frac{d\sigma
}{dr}+\left[ \frac{1}{2}\left( F_1-F_2\right) -2\alpha (\varphi
)\frac{d\varphi }{dr}\right] \frac{d\sigma }{dr} -\sigma \exp (\lambda )
\left[ \Omega ^{2}\exp (-\nu
)+4 \sigma A^{2}(\varphi )W^{\prime }(\sigma)\right], \nonumber\\
\frac{d\mu }{dr}=&F_5\equiv -\frac{g(\mu )+f(\mu )}{f^{^{\prime
}}(\mu )}\left[ \frac{1}{2}F_2+\alpha (\varphi )\frac{d\varphi
}{dr}\right] . \nonumber
\end{align}

Here the metric functions $\lambda (r)$ and $\nu (r)$, the
dilaton field $\varphi (r)$, the densities of the bosonic $\sigma
(r)$ and the fermionic $\mu (r)$ matter are unknown functions of
the radial coordinate $r \in [0, R_s] \cup (R_s, \infty) $, where
$R_s$ is the unknown radius of the star. The quantity $\Omega $
(the time-frequency of the bosonic field) is an unknown parameter
as well. The quantities  $\stackrel{\mathit{F}}{T_{n}^{n}}$,
$\stackrel{\mathit{B}}{T_{n}^{n}}$, $n = 0, 1$ (the diagonal
components of the energy-momentum tensors),
$\stackrel{\mathit{F}}{T}$, $\stackrel{\mathit{B}}{T}$ (the
corresponding traces), $\alpha(\varphi)$, $A(\varphi)$,
$V(\varphi)$, $W(\sigma)$, $f(\mu)$, and $g(\mu)$ are given
functions of their arguments, and $\gamma$ is a given constant
(see below).

Having in mind the physical assumptions, we have to set the following boundary
conditions at $r=0$:
\begin{equation}
\!\!\!\!\!\!\!\! \lambda (0)=\frac{d\varphi }{dr}(0)=\frac{d\sigma
}{dr}(0)=0,\quad \sigma (0)=\sigma _{c},\quad \mu (0)=\mu _{c},
\label{bcil} \quad \mu (R_{s})=0,
\end{equation}
\noindent where $\sigma _{c}$ and $\mu _{c}$ are the values of
density of, respectively, the bosonic and fermionic matter at the
star's centre $r=0$. The first three conditions in (\ref{bcil})
guarantee the nonsingularity of the metrics and the functions
$\lambda (r),$ $\varphi (r),$ $\sigma (r)$ at the star's centre.

In the external domain $r>R_s$ there is no fermionic matter,
i.e., one can formally suppose that the corresponding density
$\mu (r)\equiv 0$ if $x\geq R_{s}$. The fermionic part
$\stackrel{\mathit{F}} {T_{n}^{n}}$ of the energy-momentum tensor
vanishes also identically and, thus, the last equation (with
respect to the function $\mu(r)$) in (\ref{eqi}) can be removed
from the system.

As it is required by the asymptotic flatness of space-time, the boundary
conditions at the infinity are:
\begin{equation}
\nu (\infty ) = \varphi (\infty ) = \sigma (\infty )=0 \quad
\hbox{where}\quad (\cdot) (\infty) \stackrel{\rm def}{=} \lim_{r
\to \infty} (\cdot) (r). \label{bcer}
\end{equation}
We seek for a solution $\left[ \lambda (r),\nu
(r),\varphi(r),\sigma(r),\mu (r),R_{s},\Omega \right] \ $\
subjected to the nonlinear system (\ref{eqi}), satisfying the
boundary conditions (\ref{bcil}) and (\ref{bcer}). At that, we
assume the function $\mu (r)$ is continuous in the interval
$[0,R_{s}],$ while the functions $\lambda (r),$ $\nu (r)$\ are
continuous and the functions $\varphi (r),$ $\sigma (r)$ are
smooth in the whole interval $[0,\infty )$, including the unknown
internal boundary $r=R_{s}$.

The formulated boundary value problem (BVP) is a two-parametric
eigenvalue problem with respect to the quantities $R_{s}$ and
$\Omega $.

Let us emphasize that a number of methods for solving the
free-boundary problems are considered in detail in
\cite{Vab,numrec}.

Here we aim at applying the new solving method to the above
formulated problem. This method differs from the one proposed in
\cite{tomiplas} and for the governing field equations written in
the forms (\ref{eqi}) it possesses certain advantages.

\section{Method of solution}
At first we scale the variable $r$ using the Landau
transformation $ x=r/R_s, x \in [0,1]\cup (1,\infty )$ \cite{Vab}.

For given values of the parameters $R_{s}$\ and $\Omega $\ the
independent solving of the system (\ref{eqi}) in the internal
domain $x\in [0,1]$ (inside the star) requires seven boundary
conditions. At the same time, we have at disposal only six
conditions of the kind (\ref{bcil}). In order to complete the
problem, we set additionally one more parametric condition (the
value of one from among the functions $\lambda_i (x),\nu_i
(x),\varphi_i (x),\varphi^{\prime}_i (x), \sigma_i (x)$, or
$\sigma^{\prime}_i (x)$) at the point $x=1)$. Let us set for
example
\begin{equation}
\varphi _{i}(1)=\varphi _{s},\quad  \varphi _{s} -\>
\hbox{parameter.}        \label{dopusl}
\end{equation}
Obviously, the solution $\mathbf{y}_i \equiv \left
\{\lambda_i,\nu_i,\varphi_i,\sigma_i, \mu \right \}$ in the
internal domain $x\in [0,1]$ depends not only on the variable
$x,$ but it is a function of the three parameters $R_{s},\Omega
,\varphi_s$ as well.

Concerning the solution $\mathbf{y}_e \equiv \left \{ \lambda_e,
\nu_e, \varphi_e, \sigma_e \right \}$ in the external domain $x >
1$ six boundary conditions are indispensable for the solution of
the equations (\ref{eqi}). At the same time, only the three
boundary conditions (\ref{bcer}) are given. Let us consider that
the solution in the internal domain $x\in [0,1]$ is knowledged.
Then we postulate the rest three deficient conditions to be the
continuity conditions at the point $x=1.$ The first of them is
similar to condition (\ref {dopusl}) and the other two we assign
to two arbitrary functions from among $\lambda (x),$ $\nu
(x),\mathbf{\ }\varphi^{\prime} (x),$ $\sigma (x)$, and
$\sigma^{\prime} (x)$, for example
$\lambda_{e}(1)=\lambda_{i}(1);\> \varphi _{e}(1)=\varphi _{s};\>
\sigma _{e}(1)=\sigma _{i}(1).$

\medskip Let us suppose that the internal and external solutions are known.
Generally speaking, for given arbitrary values of the parameters
$R_{s},\Omega $, and $\varphi _{s}$ the continuity conditions
with respect to the functions $\nu (x), \varphi^{\prime} (x)$,
and $\sigma^{\prime} (x)$ at the point $x=1$ are not satisfied.
We choose the parameters $R_{s},\Omega $, and $\varphi _{s}$ in
such manner that the continuity conditions for the mentioned
functions to be held,
\begin{align}
\hbox{i.e.,}\hspace{2cm} \nu _{e}(1,R_{s},\Omega ,\varphi
_{s})-\nu _{i}(1,R_{s},\Omega
,\varphi _{s}) =&0,  \nonumber \\
\varphi^{\prime}_{e}(1,R_s,\Omega ,\varphi
_{s})-\varphi^{\prime}_{i}(1,R_{s},\Omega ,\varphi _{s}) =&0,
\label{cc} \\ \sigma^{\prime}_{e}(1,R_s,\Omega ,\varphi
_{s})-\sigma^{\prime}_{i}(1,R_{s},\Omega ,\varphi _{s}) =&0.
\nonumber
\end{align}
These conditions should be interpreted as three nonlinear
algebraic equations in regard to the unknown quantities
$R_s,\Omega $ and $\varphi_s$.

So, we have to solve the nonlinear BVP (\ref{eqi}) -
(\ref{dopusl}) in conjunction with the algebraic system
(\ref{cc}). The traditional technology for solving such problems,
based on the methods like shooting \cite{numrec}, has some well
known disadvantages.

In this paper we use the Continuous Analogue of Newton Method
\cite{paz} for solving the above nonlinear spectral problem which
proposes a common treatment of both differential and algebraic
problems. Detailed description of the algorithm can be found in
\cite{tomiplas}.

\section{Some numerical results}
For the purpose of illustrating we will shortly discuss some results obtained
from numerical experiments.

In the present article we consider concrete scalar-tensor model with functions
(see Section \ref{sec2})
\begin{eqnarray*}
A(\varphi )\!\!\!\!\!&=\!\!\!\!\!\!&\exp
\left(\frac{\varphi}{\sqrt{3}}\right), \quad V(\varphi )=
\frac{3}{2}[1-A^{2}(\varphi )]^2,\quad W(\sigma)
=-\frac{1}{2}\left( \sigma ^{2}+\frac{1}{2}\Lambda \sigma^4
\right),\\
&& f(\mu ) =\frac{1}{8}\left[ (2\mu -3)\sqrt{\mu +\mu ^{2}}+3\ln
\left( \sqrt{\mu } + \sqrt{1 + \mu }\right) \right],\\
&& g(\mu ) =\frac{1}{8}\left[ (6\mu +3)\sqrt{\mu +\mu ^{2}}-3\ln
\left( \sqrt{\mu }+\sqrt{1+\mu }\right) \right].
\end{eqnarray*}

The quantities $b$ and $\Lambda$ are given parameters. The
functions $f(\mu)$ and $g(\mu)$ represent in parametric form the
equation of state of the non-interacting neutron gas, while the
function $W(\sigma)$ describes the boson field with quadratic
self-interaction.

The behaviour of the calculated eigenfunctions $\sigma(x)$,
$\varphi(x)$, $\nu(x)$, and $\mu(x)$ is typical for wide range of
the parameters and was discussed in detail in \cite{tomiplas}.

From a physical point of view, it is important to know the mass
of the boson-fermion star and the total number of particles
(bosons and fermions) making up the star.

The dimensionless star mass can be calculated by the formula:
$$ M= \int\limits_{0}^{\infty} r^2 \left[\stackrel{\mathit{B}}{T_{0}^{0}}+
\stackrel{\mathit{F}}{T_{0}^{0}} +
\exp(-\lambda)\left({\frac{d\varphi }{dr}} \right)^2 +
{\frac{\gamma^2}{2}}V(\varphi) \right] dr.
$$
The dimensionless rest mass of the bosons (total number of bosons
times the boson mass) and the dimensionless rest mass of the
fermions are given by:
$$
M_{RB}= \Omega \int\limits_0^{\infty} r^2 A^2(\varphi)
\exp\left({\frac{\lambda - \nu }{2}}\right)\sigma^2 dr, \qquad
M_{RF}=b\int\limits_{0}^{\infty }r^{2}A^{3}(\varphi )
\exp\left({\frac{\lambda }{2}}\right) n(\mu ) dr,$$ where
$n(\mu)$ is the density of the fermions. In the case we have
$n(\mu)=\mu ^{\frac{3}{2}}(x)$.

The dependencies of the star mass $M$ and the rest mass of
fermions $M_{RF}$ on the central value $\mu_{c}$ of the function
$\mu(x)$ are shown in the configuration diagram on Fig. 1 for
$\Lambda =0$, $\gamma =0.1$, $b=1$ and $\sigma _{c}=0.002$. It
should be pointed that for such small central value $\sigma_{c}$
we actually have a ``pure" fermionic star. On the figure, it is
seen that from small enough values of $\mu_{c}$, in the vicinity
of the the peak the rest mass is greater than the total mass of
the star, which means that the star is potentially stable.

\begin{figure}
\centerline{\psfig{figure=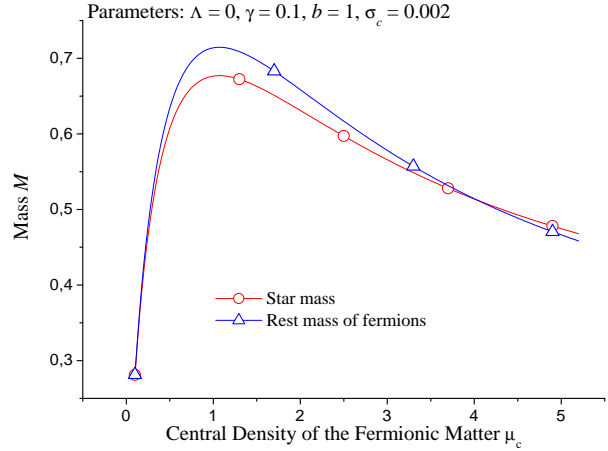,width=4.in}} \caption{\small
The star mass $M$ and the rest fermion mass $M_{RF}$ as functions
of the central value $\mu_c$.}
\end{figure}
\begin{figure}
\centerline{\psfig{figure=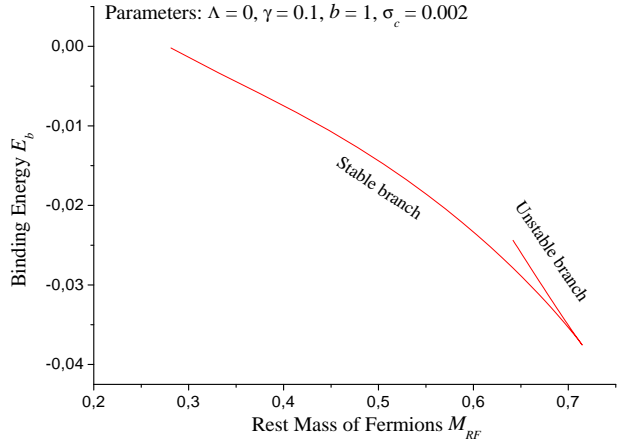,width=4.in}} \caption{\small The
binding energy $E$ versus the rest fermion mass $M_{RF}$.}
\label{bind}
\end{figure}

On Fig. 2 the binding energy of the star $E_{b}=M-M_{RB}-M_{RF}$
is drawn against the rest mass of fermions $M_{RF}$ for $\Lambda
=0$, $\gamma =0.1$, $b=1$ and $\sigma _{c}=0.002$. Fig. 2 is
actually a bifurcation diagram. By increasing the central value
of the function $\mu(x)$ a cusp appears. The presence of the cusp
shows that the stability of the star changes~- one perturbation
mode develops instability. Beyond the cusp, the star is unstable
and may collapse, eventually forming a black hole. The
corresponding physical results for pure boson stars are
considered in our recent paper \cite{FYBT2}.

\bigskip

{\bf{Acknowledgements.}} SSY is grateful to the organizers of the
$4^{th}$ General Conference of the Balkan Physical Union and
especially to Prof. M. Mateev for the opportunity to give this
talk and for financial support. TLB and MDT thank Prof. I.V.
Puzynin (JINR, Dubna, Russia) for efficient discussions.

\end{document}